\documentclass[a4paper,12pt]{article}

\begin{document}
\title{ On a connection between gravitation and quantum chromodynamics.}
\author{ J. A. Martins Sim\~oes,\thanks{E-mail: simoes@.if.ufrj.br}\\ Instituto
de F\'{\i}sica,\\
Universidade Federal do Rio de Janeiro, RJ, Brazil}
\maketitle
\begin{abstract}
\par
In this paper we point out a remarkable numerical relation between the large scale parameters of gravitation $ G_N$ and $H_0$ and the two fundamental parameters of quantum cromodynamics $ \alpha_{QCD}$ and $ \Lambda_{QCD}$.  With this  remark we make the conjecture that Newton's gravitational constant could show an energy dependence analogous to $ \alpha_{QCD}$.

\end{abstract}
\eject

 In 1938 Dirac found a numerical relation between the large scale parameters of gravitation, Newton's gravitational constant and Hubble's constant,$G_N$ and $H_0$ respectively, and some of the small scale parameters of elementary particles, the Planck constant, the proton mass and the electron charge. This relation is given by

\begin{equation}
  {{G_N \over H_0 }\over {h^3 \epsilon _0 \over m_p^3 e^2}}=1.3
\end{equation}
and is numerically satisfied at a high degree of precision.
\par
 The cosmology that followed this relation suggested that the gravitational parameter $G_N$ had a time dependence. As no such effect was experimentally found \cite {BER}, this connection between large and small scale parameters was abandoned.
\par
However, Dirac's original argument has a physical point that deserves more attention. Any connection between parameters at different scales is an indication of a deeper physical relation of some  fundamental principles. So we have taken Dirac's program from a more modern point of view. 
\par
Our first modification in Dirac's argument is that the proton mass is no longer a fundamental mass scale. According to the standard model of elementary particles, the fundamental mass scales are the vacuum symmetry breaking parameter $ v_{Fermi}$ and the QCD scale $ \Lambda _{QCD}$. The second modification is that as $G_N$ is a coupling constant, it must be proportional to $g_x ^2$, where "x" means some of the standard model couplings, and not to the inverse power of the electron charge, as in Eq. 1. 
\par
If we now add Planck's constant and the speed of light, then a pure dimensional analysis leads us to the equation
\begin{equation}
  {G_N \over H_0 }={h^2 \over c}{{g_{QCD}^2} \over {\Lambda_{QCD}^3}}
\end{equation}
\par 
Using the last 2004 Particle Data Group \cite{PDG} numerical values for physical constants and $h=0.71$ in Hubble's constant, $\alpha_{QCD}=0.12$ and $\Lambda_{QCD}=220 MeV/c^2$, we have the result that the two terms differ by a factor 0.8, which is very close to one. The large  difference on the  scales of physical phenomena involved in this new version of Dirac's relation strongly suggests  some fundamental connection between gravitation and QCD. Any other numerical relation between the standard model couplings and fundamental mass scales will not satisfy relations like Eq. 2. The consequences of this equation can be pursued in at least three ways.
\par
There is an experimental evidence \cite{WEB} that the electromagnetic fine-structure constant has a small variation from the value it had some billions of years ago. After this result, Calmet and Fritzsch  \cite{FRI} proposed that the QCD parameters could also have changed during this cosmic time scale. Our Eq. 2 predicts that any time variation in the QCD parameters will imply time variations in the cosmological parameters and vice-versa.
\par
The other important consequence of Eq. 2 concerns the energy dependence of $G_N$ and $\alpha _{QCD}$. In the QCD case, the behavior of $\alpha_{QCD}(Q^2)$ is well known and leads to asymptotic freedom. A $Q^2$ dependence of 
$G_N$ is  not a surprise in field theory. However, the non-renormalization of gravitation prevents an analysis similar to the QCD case. In more recent years the effective field theory method has shown \cite{DON} interesting results in calculating corrections for some cases of gravitational interactions. For example, the leading long distance quantum corrections of a system of two interacting charged masses can be computed. But the general small distance behavior of $G_N$ is not known. If we admit that the connection given by Eq. 2 is to be valid in some energy range and that the energy dependence is only in the coupling terms, then we would have the relation
\par
\begin{equation}
  G_N(Q^2) \propto \alpha_{QCD}(Q^2)
\end{equation}
 This relation must be viewed as a conjecture following from Eq. 2 and not as a demonstration. We call it the " Leme conjecture" \cite{LEM}. We expect that this conjecture holds in the energy range of validity of Eq. 2  and that it will imply in logarithm corrections to gravitational phenomena at different energy scales. But as the  formal proof of this conjecture is still lacking , we can't fully justify its extention to the asymptotic small distance region. 
\par
Finally, the main question that Eq. 2 raises is on the possible physical scenario from which one could naturally derive this relation. The answer to this question could clarify the small and large scale  behavior of gravitation and QCD. In particular, if our conjecture is confirmed, it would mean that the hadrons and the universe are  related systems.
\vskip 1cm

{\it Acknowledgments:}  We thanks T.J. Martins Sim\~oes for suggestions concerning reference 5. This work was partially supported by the following Brazilian agencies: CNPq and FAPERJ.


\begin{thebibliography}{ABC}
\bibitem{LEM} The name is after a famous beach in Rio, where the author lives.
\end{thebibliography}
\end{document}